\documentclass[prb,aps]{revtex4}
\usepackage{graphicx}
\usepackage{dcolumn}
\usepackage{bm}
\usepackage{amsmath}
\usepackage{amssymb}

\newcommand{\down}{\downarrow}

\begin{document}

\title{Fermionic trimers in spin-dependent optical lattices}

\author{Giuliano Orso$^1$}
\author{Evgeni Burovski$^2$}
\author{Thierry Jolicoeur$^2$}

\affiliation{$^1$Laboratoire Mat\' eriaux et Ph\' enom\`enes Quantiques,
Universit\' e Paris Diderot-Paris 7 and CNRS, UMR 7162, 75205 Paris Cedex 13, France}
\affiliation{
$^2$LPTMS, CNRS and Universit\'e Paris-Sud, UMR8626, Bat. 100, 91405 Orsay, France
}

\date{October 15th, 2010}
\begin{abstract}
We investigate the formation of three-body bound states (trimers) in two-component Fermi gases confined in 
one dimensional  optical lattice with spin-dependent tunneling rates. The binding energy and the effective 
mass of the trimer are obtained from the solution of the  Mattis integral equation generalized to the case 
of unequal Bloch masses. 
We show that this equation admits multiple solutions corresponding to excited bound states, which are only 
stable for large mass asymmetry.     
\end{abstract}
\maketitle

\section{Introduction}
\label{intro}

The advent of Feshbach resonances and optical lattices has
caused a major revolution in the field of ultra-cold atoms \cite{RMP2008}. 
Few-body physics is currently playing an important and intriguing role. First, 
bound states of few interacting atoms can be produced and studied experimentally in 
a controlled way, providing a direct test to fundamental quantum-mechanics. Recent examples are  
the observation of repulsively bound pairs of bosonic atoms in an optical lattice 
\cite{repulsivelybound} and the detection of an Efimov-like trimer  \cite{efimov} of ${}^{133}$Cs atoms.

Second, optical lattices can considerably affect the many-body scenario by 
modifying the two-body scattering properties of interacting atoms \cite{fedichev,Tc2005}.
The connection between few- and many-body physics is
particularly interesting in one dimensional (1D) systems, where the strongly interacting 
regime typically occurs at \textsl{low} density. 
After the first observation \cite{esslinger} of confinement-induced molecules in attractive 
Fermi gases, experimentalists are now addressing the rich many-body scenario predicted in these systems. 

In this context, a very recent experiment \cite{Hulet2010}  performed at Rice University 
investigated the properties of a one dimensional trapped  Fermi gases with attractive 
interaction and a finite spin polarization, 
verifying the two-shell structure of the density profiles predicted in 
Refs.\cite{Orso2007,Drummond2007} on the basis of Bethe-Ansatz calculations. 
In particular the partially polarized core of the gas is expected to be a superfluid 
of the Fulde-Ferrell-Larkin-Ovchinnikov (FFLO)  type \cite{FFLO,yang}, as confirmed 
by exact numerical simulations \cite{Feiguin}. 

Motivated by the strong interest in this field, in Refs.\cite{bosonization,PRL2010} 
we investigated the ground state properties of a two-component Fermi gas confined in a 
1D lattice with spin-dependent tunneling rates. The system is
described by the asymmetric Fermi-Hubbard model ~\cite{GiamarchiCazalillaHo2005,Batrouni2009,DasSarma2009}:
\begin{equation}
H = -\sum_{ i \sigma} t_\sigma \left(
c^\dagger_{i,\sigma} c_{i+1,\sigma} +h.c. \right) +
U\sum_{i}\hat{n}_{i\uparrow}\hat{n}_{i\downarrow} \; ,
\label{aHubb}
\end{equation}
where $U<0$ is the on-site attraction and $t_\sigma$ are the spin-dependent
tunneling rates. Here $c_{i\sigma}$ annihilates a fermion with spin $\sigma$
at site $i$  and $\hat{n}_{i\sigma}$ is the local density. 
The Hamiltonian (\ref{aHubb}) can also describe mixtures of two different atomic
species, like mixtures of $^6$Li and $^{40}$K near a heteronuclear Feshbach resonance 
\cite{Wille2008,dieckmann,unequal}. 

For equal tunneling rates, $t_\down=t_\uparrow$, the exact Bethe Ansatz solution  
of the model (\ref{aHubb}) shows that $n$-body bound states with $n>2$ are  generally forbidden \cite{takahashi}. 
In Ref.\cite{PRL2010} we showed that in the presence of unequal tunneling rates, there is formation 
of trimers made of two heavy ($\down$) fermions and one light
 ($\uparrow$) fermion. These states are reminescent of the trions recently observed \cite{trion} in semiconductors. 
 Then a DMRG calculation shows that 
 trimers are responsible for the appearance of a new gapped phase for finite and commensurate densities 
$(n_\down =2 n_\uparrow)$, which is characterized by exponential suppression of both single-particle 
and superfluid FFLO correlations. 

The purpose of this paper is to present a more thorough derivation of the three-body calculations 
 outlined in Ref. \cite{PRL2010}. We also discuss new and unexpected results for the excited bound states.
The three-body problem of interacting atoms with unequal mass has already attracted a lot of 
attention \cite{P,KML,KM,NT,LTWP,KMS,HHP,PP}. We will make contact with the corresponding results for the continuum model obtained in Ref.\cite{KMS}.

The article is organized as follows. In section \ref{sec2} we present a self-contained derivation of
 the Mattis integral equation \cite{mattis} for the three-body problem, generalized to the case of
 unequal masses. In section \ref{sec3} we calculate the ground state solution of such equation, 
corresponding to the trimer state with the highest binding energy. In section \ref{sec4} we investigate 
the excited bound states solutions of the integral equation. Finally we give our conclusions in section \ref{sec5}.

Before continuing, we would like to mention that three-body bound states (though of different nature) also occur
in the 1D Bose-Hubbard model, as recently investigated in Ref.\cite{valiente10,keilmann09}, 
as well as in Fermi gases with three or more spin components \cite{azaria09,kantian09,luescher10}.

\section{Integral equation for trimers}
\label{sec2}

In this section we consider two $\down$-fermions interacting with one $\uparrow$-fermion, 
as described by the Hamiltonian (\ref{aHubb}). We map the Schr\"{o}dinger equation of the three particles 
into an integral equation which we then solve both analytically and numerically.  
The Schr\"{o}dinger equation  in momentum space takes the form
\begin{equation}
(\epsilon_\down (k_1)+ \epsilon_\down (k_2) + \epsilon_\uparrow (k_3)-E)      
\psi(k_1,k_2,k_3) + U \int \frac{dp}{2 \pi} \psi(p,k_2,k_1+k_3-p) +
U \int \frac{dp}{2 \pi} \psi(k_1,p,k_2+k_3-p) =0,
\label{der1}
\end{equation}
where $\epsilon_\sigma(k)=2 t_\sigma (1-\cos k)$ the energy dispersions  of the two
components and the integration over quasi-momenta is restricted to $[-\pi,\pi]$. We see from Eq.(\ref{der1}) 
that the total quasi-momentum $P=k_1+k_2+k_3$ is a conserved
quantity associated to the discrete traslational invariance of the lattice model \cite{noteP}. Introducing the function 
\begin{equation}
\label{der2}
A(k,P)=\int_{-\pi}^\pi \frac{dp}{2\pi}\psi(p,k,P-p-k), 
\end{equation}
and taking into account that the wavefunction is antisymmetric under exchange of the two $\downarrow$ fermions,  
$\psi(k_1,p,P-k_1-p)=-\psi(p,k_1,P-k_1-p)$,
we can rewrite Eq.(\ref{der1}) as 
\begin{equation}
\label{der3}
\psi(k_1,k_2,k_3)=-U\frac{A(k_2,P)-A(k_1,P)}{\epsilon_\down (k_1)+ \epsilon_\down (k_2) + \epsilon_\uparrow (k_3)-E},
\end{equation}
which can be seen as a self consistent equation for the function $A(k,P)$. Inserting Eq.(\ref{der3}) into
 Eq.(\ref{der2}) we obtain
\begin{equation}
\label{der4}
A(k,P)=-U \int_{-\pi}^{\pi} \frac{dp}{2\pi}\frac{A(k,P)-A(p,P)}{{\mathcal E}(k,q,P)-E},
\end{equation}
where ${\mathcal E}(k,q,P)=\epsilon_\downarrow (k)+\epsilon_\downarrow (q) +
\epsilon_\uparrow(P-k-q)$ is the total energy dispersion. Bringing the term proportional to $A(k,P)$ in 
Eq.(\ref{der4}) to the right hand side, we get
\begin{equation} 
\label{der5}
A(k,P)(1+U I_E(k,P))=U \int_{-\pi}^{\pi} \frac{dp}{2\pi} \frac{A(p,P)}{{\mathcal E}(k,q,P)-E},
\end{equation}
where the integral $I_E(k,P)$ is defined by
\begin{equation}
\label{I_E}
I_E(k,P)=\int \frac{dp}{2\pi}\, \frac{1}{{\mathcal E}(k,p,P) -E} \; .
\end{equation}
By setting $z=e^{i p}$ and using the residue theorem for complex functions, we find
\begin{equation}
\label{I_Ean}
I_E(k,P)=\frac{1}{\sqrt{(\epsilon_\downarrow(k)+2 t_\downarrow+2-E)^2-4-4 t_\downarrow^2 -8 t_\downarrow \cos(P-k)}}\; .
\end{equation}
The function $I_E(k,P)$ appears already in the solution of the two-body problem in a lattice \cite{twobody}. 
In particular 
the condition  $1+ U I_E(k=0,P=0)=0$ yields the binding energy $E_\mathrm{pair}^b=-E$ of a pair of up and 
down fermions. From Eq.(\ref{I_Ean}) one finds $E_\mathrm{pair}^b=-2(1+t_\downarrow)+\sqrt{U^2+4(1+t_\downarrow)^2}$ 
\cite{moelmer}.

Introducing $R_E(k,P)=(1+U I_E (k,P))^{1/2}$ and the function $f(k,P)=A(k,P)R_E(k,P) $, Eq.(\ref{der5}) 
takes the form of a homogeneous integral equation ~\cite{mattis}
\begin{equation}
\label{int}
f(k,P)=\int_{-\pi}^{\pi} \frac{dq}{2\pi}\,
\frac{U f(q,P)}{R_E(k,P)R_E(q,P)\left[{\mathcal E}(k,q,P)
-E\right]} \; ,
\end{equation}
 whose solution yields the energy $E$ of the three-body system. 
 We are interested here on bound states solutions corresponding to energy $E< -E_\textrm{pair}^{b}$.  


 It is important to notice that for zero total quasi-momentum $(P = 0$), 
 the function $f(k,P=0)$ in Eq.(\ref{int}) must be odd, namely $f(-k,0)=-f(k,0)$. 
 This comes from the fact that in this limit the function (\ref{der2})
coincide with the Fourier transform of the real space wave-function when two interacting particles are at the same lattice site:
\begin{equation}
\label{phys}
\psi(n_1,n_2,n_1)=\int_{-\pi}^{\pi} \frac{dk}{2\pi} e^{ik(n_2-n_1)}A(k,P=0),
\end{equation}
where $\psi$ is a zero quasi-momentum state.
Since $\psi(n_1,n_2,n_1)=-\psi(n_2,n_1,n_1)$, we see from Eq.(\ref{phys}) that
$A(k,P=0)$ is an odd function of the quasi-momentum. Since $R_E(-k,P=0)=R_E(k,P=0)$, we 
conclude that the function $f(k,0)$ is also odd.  This property will be used systematically below to obtain 
our analytical results.

In the following we first discuss the properties of the three-body bound state
with the lowest energy $E$ (or, equivalently, with the largest binding energy). 
Excited bound states solutions will be discussed next. To simplify the notation, 
from now on, we fix the energy scale by set $t_\uparrow=1$ in Eq.(\ref{int}).

\section{Ground state solution}
\label{sec3}

\subsection{Binding energy}
\begin{figure}
\begin{center}
\includegraphics[width=9cm]{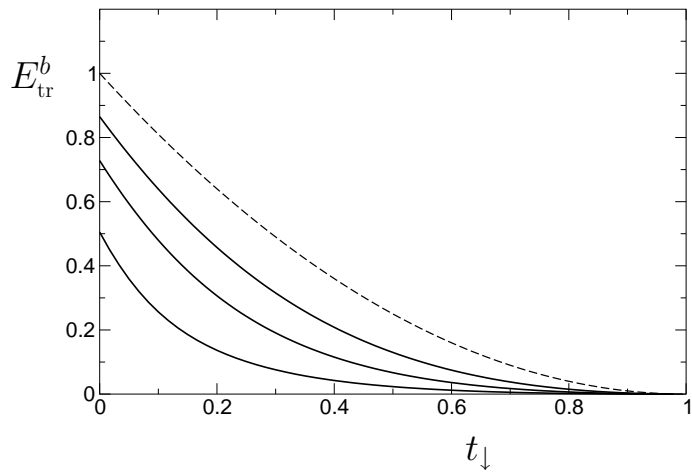}
\caption{Binding energy of the ground state trimer as a function of the 
tunneling rate $t_\down$ for different values of the interaction $U=-2 \textrm{(bottom curve)},-4,-8$. 
The asymptotic curve 
[see Eq.(\ref{strong})] in the strong coupling limit $U=-\infty$ is shown with the dashed line.}
\label{fig1}
\vspace{1cm}
\end{center}
\end{figure}

Equation \ (\ref{int}) can be considered
as an eigenvalue problem $\bf K_E \cdot \bf f= \lambda \bf f$, where the
energy $E$  is fixed by the constraint $\lambda=1$. We solve this equation numerically for 
zero quasi-momentum $P=0$. The binding energy $E_\mathrm{tr}^b$ of the trimer is related to the total energy
$E$ by $-E=E_\mathrm{pair}^b + E_\mathrm{tr}^b$. 
In Fig.\ \ref{fig1} we plot the binding energy of the trimer as
a function of the mass asymmetry $t_\downarrow$ for increasing values of the
attraction $U$.

We see that $E_\mathrm{tr}^b$ vanishes at the symmetric point
$t_\downarrow=1$ for any values of the interaction $U$, in agreement with the Bethe Ansatz 
solution~\cite{takahashi}. For $t_\down >1$ no bound state solution
has been found.
This result can be understood by noticing that when the two heavy particles approach each 
other, the light fermion can hop between the two
without loss of potential energy. Therefore, if $t_\downarrow < 1$, the energy gain to 
delocalize the light particle overcompensates the energy cost to localize the heavy 
fermions, and the trimer state is bound.

As the mass asymmetry increases the binding energy also increases until it saturates  at 
$t_\down \rightarrow 0$, where the effective mass of the heavy fermions becomes infinite. 
In this limit  the function (\ref{I_Ean}) reduces to a constant 
$I_E(k)=1/\sqrt{E(E-4)}$, implying that $R_E(q)=(1+U/\sqrt{E(E-4)})^{1/2}=R_E $
in Eq.\ (\ref{int}) is also constant. 
By changing the integration variable to $q^\prime=q+k$ and omitting the prime index, the latter takes the form
\begin{equation}
\label{int_tD0}
f(k)= \frac{U}{R_E^2} \int_{-\pi}^{\pi} \frac{dq}{2\pi}\,
 \frac{f(q-k)}{2(1-\cos q)-E} \; .
\end{equation}

\begin{figure}
\begin{center}
\includegraphics[width=9.2cm]{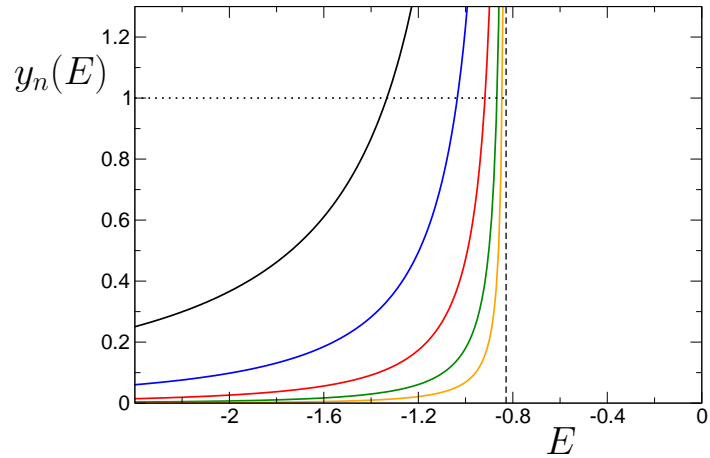}
\caption{(Color online) Energy dependence of the function $y_n(E)$  [see Eq.(\ref{y_n})] 
plotted for $U=-2$ and for different values of the index $n=1 \mathrm{(
top \,  curve)}, 2,3,4,5$. The dashed vertical line corresponds to the threshold energy 
$-E_b^\textrm{pair}=2-\sqrt{U^2+4}=-0.828$  for the pair bound state for $t_\down=0$. The 
energy of the bound state is found from the solution of the equation $1=y_n(E)$. The ground 
state corresponds to $n=1$, where $E=-U^2/(1-U)=-1.333$ [see Eq.(\ref{falikov})].  }
\label{fig2}
\vspace{1cm}
\end{center}
\end{figure}

Taking into account that $f(k)$ is an odd function, we can write the general solution as 
$f(k)=\sum_{n=1}^{n=\infty} a_n \sin (nq)$. Substituting this into Eq.(\ref{int_tD0}),
we find that the different harmonics decouple, implying that the solution is of the form 
$f_n(k)=\sin (nk)$. By inserting this into Eq.(\ref{int_tD0}), the latter 
reduces to $1=y_n(E)$, where
\begin{equation}
\label{y_n}
y_n(E)=\frac{U}{R_E^2} \int_{-\pi}^{\pi} \frac{dq}{2\pi}\,
 \frac{-\cos(n q)}{2(1-\cos q)-E} \; .
\end{equation}

The function $y_n(E)$ is plotted in Fig.\ref{fig2} for $U=-2$ and for increasing values of 
$n$, starting from $n=1$ (top curve). The dashed vertical line corresponds to $E=-E_\mathrm{pair}^b=2-\sqrt{U^2+4}$. 
We see that there is an infinite number of three-body bound state solutions that for large 
$n$ accumulates near the pair energy.  
The ground state solution corresponds to $n=1$, namely $f(k)=\sin k$. 
By using the formula $ \int_{-\pi}^{\pi}  \cos q/(a-\cos q) dq=2\pi(a/\sqrt{a^2-1}-1)$ 
valid for $a>1$, as well as the explicit expression for $R_E$, from  Eq.(\ref{y_n}) we obtain 
\begin{equation}
1=\frac{-U}{2}\frac{(2-E-\sqrt{E(E-4)}}{\sqrt{E(E-4}+U},
\end{equation}
yielding $E=-U^2/(1-U)$. Therefore the trimer binding energy at $t_\down=0$
is given by \cite{PRL2010}
\begin{equation}\label{falikov}
 E_\mathrm{tr}^b(t_\down= 0)=\frac{U^2}{1-U}+2-\sqrt{U^2+4}\;,
\end{equation}
in agreement with our numerical results in Fig.\ref{fig1}.
In this limit the problem is very simple since we are dealing with two fixed scatterer
and only one mobile particle~: this point of view will be used in detail in section 4.

Let us now discuss the dependence of the binding energy on the interaction $U$.
Clearly, the binding energy increases as $U$ increases. However, differently from the pair 
binding energy, which diverges for infinite attraction,   $E_\mathrm{tr}^b$ saturates to 
a finite value shown in 
Fig.\ \ref{fig1} with dashed line. 
In this strong coupling regime, corresponding to $|E| \sim |U| \gg 1$, we can use the  
expansion $({\mathcal E}-E)^{-1} \simeq-1/E-{\mathcal E}/E^2$ in Eq.(\ref{int}). The 
first term gives no contribution due to symmetry considerations, whereas the second term yields
\begin{equation}
\label{intU}
f(k)=\int_{-\pi}^{\pi} \frac{dq}{2\pi}\,
\frac{U \cos(q+k) f(q)}{E^2 R_E(k)R_E(q)} \; .
\end{equation}
By using the formula $\cos(q+k)=\cos q \cos k -\sin q \sin k$ in Eq.(\ref{intU}), we 
immediately see that the solution must be of the form $f(k)=\sin k/R_E(k)$. A direct substitution then yields
\begin{equation}\label{strong1}
1=\frac{-2U}{E^2} \int_{-\pi}^{\pi} \frac{dq}{2\pi} \frac{\sin^2 q}{R_E(q)^2}=
\frac{-2U}{E^2} \int_{-\pi}^{\pi} \frac{dq}{2\pi} \frac{\sin^2 q}{1-U/E-2U(1+2 t_\down-t_\down \cos q)/E^2},
\end{equation}
where in the second equality we have made use of the asymptotic expansion
of $I_E(q)$ from Eq.(\ref{I_Ean}). Next, we write $E=U+ \alpha$ in Eq.(\ref{strong1})
and take the limit $U \rightarrow -\infty$ assuming the energy shift $\alpha \ll |U|$. This yields 
\begin{equation}
1=2 \int_{-\pi}^{\pi} \frac{dq}{2\pi} \frac{\sin^2 q}{-\alpha+2(1+2 t_\down)-2t_\down \cos q},
\end{equation}
which can be solved analytically to obtain the shift $\alpha$. By using the formula
$ \int_{-\pi}^{\pi}  \sin^2 q/(a-\cos q) dq=2\pi(a-\sqrt{a^2-1})$ valid for $a>1$, we find
$\alpha=4 t_\down-t_\down^2+1$. Making use of the strong coupling expansion  
$E_\mathrm{tr}^b(U\rightarrow -\infty)\simeq
-U-2-2 t_\down$, we finally obtain \cite{PRL2010}
\begin{equation}\label{strong}
E_\mathrm{tr}^b(U=-\infty)=(t_\downarrow-1)^2,
\end{equation}
showing explicitly that the binding energy of the trimer remains finite even in the
strongly interacting regime.

\begin{figure}
\begin{center}
\includegraphics[width=9.2cm]{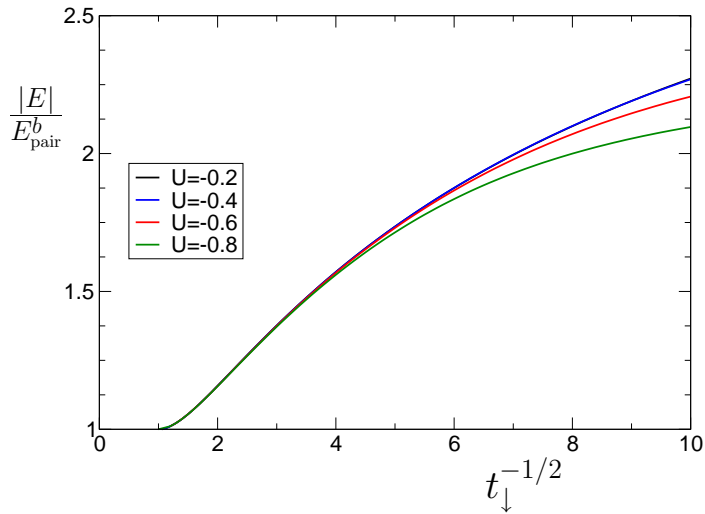}
\caption{(Color online) Ratio between the trimer and pair energies as a function of $t_\downarrow^{-1/2}$ calculated for decreasing values of the interaction $U$. 
Notice that for $|U|\leq 0.4$ the curves are on top of each other signaling convergence to the continuum limit.  We then recover the results for
 the continuum model obtained in Ref.\cite{KMS} (see their Fig.1, left panel) 
 with the (Bloch) masses given by $m_\sigma=1/2 t_\sigma$.   }
\label{fig3}
\vspace{1cm}
\end{center}
\end{figure}

It is also interesting to consider the opposite limit of weak interaction, namely 
$|U| \ll t_\down,t_\uparrow$. In this case
only the states at the bottom of the band are important and we can 
approximate the tight-binding dispersions  with the quadratic expansions $\epsilon_\sigma (k) \simeq k^2/2 m_\sigma$, where $m_\sigma=1/2 t_\sigma$ are the 
related Bloch masses. This corresponds to the continuum model studied in Ref.\cite{KMS}. 
The convergence to the continuum result is studied in Fig.\ref{fig3} where 
we plot the ratio $|E|/E_\textrm{pair}^b$ between the trimer and pair energies 
as a function of $t_\downarrow^{-1/2}$ for decreasing values of the interaction 
strength $|U|$. We see that for 
the range of mass asymmetry considered here all curves with $|U|\lesssim 0.4$
are on top of each other and coincide with the continuum prediction of Ref.\cite{KMS}.
Clearly deviations from the continuum limit are stronger for large mass asymmetry,
where the condition $- U\ll t_\down$ becomes more stringent.

\subsection{Effective mass}
\begin{figure}
\begin{center}
\includegraphics[width=9cm]{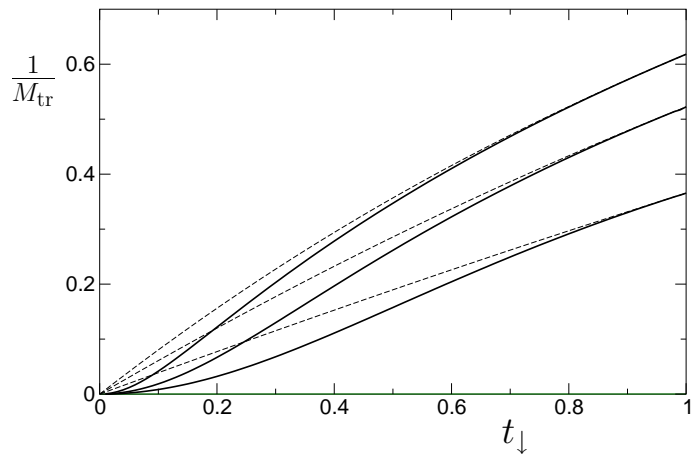}
\caption{Inverse effective mass of the trimer as a a function of the tunneling rate $t_\down$ 
for different values of the interaction $U=-2 \textrm{(upper solid curve)},-4,-8$. The 
corresponding asymptotic curves 
[see Eq.(\ref{totalmass})] in the symmetric limit $t_\down=1$ are shown with the dashed lines.}
\vspace{1cm}
\label{fig4}
\end{center}
\end{figure}

It is also interesting to  discuss the effective mass $M_\mathrm{tr}$ of the trimer. 
The latter is related to the energy dispersion $E(P)$ of the trimer by 
$1/M_\mathrm{tr}=\partial^2 E/\partial P^2$ evaluated at
$P=0$. We replace the derivative by a finite difference that we evaluate numerically. 
The result for the inverse effective mass is plotted in the inset of Fig.\
\ref{fig4}  as a function of the hopping rate $t_\downarrow$ and for
different values of the attraction strength. We see that the trimer becomes heavier as 
$t_\down$ decreases or $|U|$ increases. 

At the symmetric point $t_\down=1$,
where the trimer disappears, the effective mass  must coincide with the sum of the masses of its constituents. 
The effective mass of the pair can be calculated by the same formula, starting from the 
relation $1+U I_E(0)=0$, where the function $I_E$ in Eq.(\ref{I_Ean}) is evaluated at 
finite total quasi-momentum $P \neq 0$. This yields the 
energy dispersion $E(P)=2+2t_\down-\sqrt{U^2+ 4 +4 t_\down^2+8 t_\down \cos P }$
for the pair, from which we obtain $1/M_\textrm{pair}=4 t_\down/\sqrt{4 +8 t_\down +4 t_\down^2+U^2}$. 
Since the effective mass of the heavy fermion
is simply given by $1/(2 t_\down)$, we find that the total mass $M_\mathrm{tot}$ of the constituents is given by
\begin{equation}\label{totalmass}
M_\mathrm{tot}=(\sqrt{4(t_\downarrow+1)^2+U^2}+2)/4t_\downarrow, 
\end{equation}
which is shown in Fig.\ref{fig4}  with dashed lines.

\section{Excited bound states}
\label{sec4}

So far we have discussed the ground state solution corresponding to the bound state with 
the lowest energy  (or, equivalently, with the largest binding energy). As mentioned in 
Section \ref{sec3} Eq.(\ref{int}) admits other solutions  corresponding to excited bound 
states, namely states with energy $E=E_n$ satisfying  $E_1<E_n <-E_b^\textrm{pair}$, 
where $E=E_1$ is the ground state energy. 

\begin{figure}
\begin{center}
\includegraphics[width=9cm]{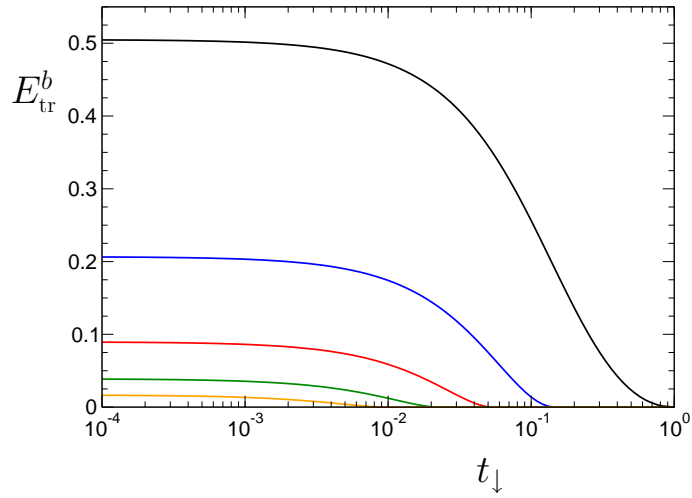}
\caption{(Color online) Binding energies of the ground state (upper curve) and the first four 
excited three-body bound states as a function of the hopping ratio $t_\downarrow$ and for $U=-2$. 
The excited bound states are only stable for large mass asymmetries $(t_\down \ll 1)$.}
\label{fig5}
\vspace{1cm}
\end{center}
\end{figure}

In Fig.\ref{fig5} we plot binding energy versus tunneling rate $t_\down$ of the 
ground state (top curve) and the first four excited bound states calculated for $U=-2$. 
We see that these excited states  are only stable for sufficiently large mass asymmetries, 
corresponding to $t_\down \ll 1 $. In the limit $t_\down=0$, Eq.(\ref{int}) admits an infinite 
number of solutions of the form $f_n(k)=\sin(n k)$. The corresponding energy levels $E=E_n$ 
are obtained  from the condition 
 $1=y_n(E)$, where the function $y_n(E)$ is defined in Eq.(\ref{y_n}). 
To understand this fact we notice that $R_E(k)=R_E$ is constant for $t_\down =0$, so 
  from Eq.(\ref{phys}) we find that 
 \begin{equation}\label{1body}
\psi_n(n_1,n_2,n_1) = \frac{A_n}{2 i R_E }\left[\delta_{n_2-n_1,-n}-\delta_{n_2-n_1,n}\right ],
\end{equation}
where $A_n$ is a normalization factor. Equation (\ref{1body}) shows that the index $n$ 
corresponds to the physical distance between  the two (infinitely) heavy fermions. 
One is therefore left with the problem of a single light fermion in the presence of two
\textsl{static contact}  potentials separated by a distance $d=n$. An explicit solution 
shows that the ground state is bound for any distance $d$, and the corresponding energy 
is given by $E=E_n$.
It should be noticed that the case $n=0$, corresponding to two heavy fermions at the same 
site, is forbidden in our three-body problem by the Pauli exclusion principle [the 
wave-function (\ref{1body}) vanishes]. As a consequence the ground state solution 
corresponds to $n=1$, where the heavy particles are nearest neighbor.

In the presence of a finite tunneling rates, $t_\down \neq 0$, the heavy fermions  
delocalize more and more  affecting significantly the stability of these excited bound states, 
as shown in Fig.\ref{fig5}.

\begin{figure}
\begin{center}
\includegraphics[width=9cm]{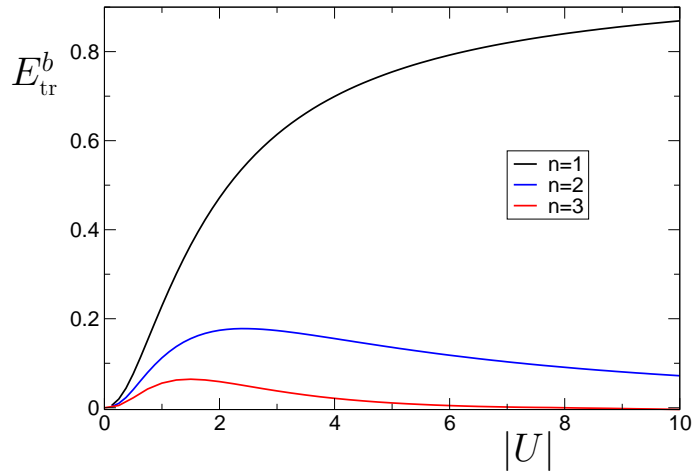}
\caption{(Color online) Binding energies of the ground state (upper curve) and the first two 
excited three-body bound states as a function of the interaction $|U|$ for fixed $t_\downarrow=0.01$. 
Notice the non-monotonic dependence of the binding energy for the excited states 
$n>1$.}
\label{fig6}
\vspace{1cm}
\end{center}
\end{figure}

Finally, it is interesting to discuss the behavior of the binding energy of the excited states 
as a function of the interaction strength. This is shown in Fig.\ref{fig6} for a fixed value 
$t_\down=0.01$ of the tunneling rates. We see that differently from the ground state solution,
the binding energy for $n>1$ exhibits a non-monotonic behavior as a function of $U$, with 
a maximum around $U \sim - 2$. This comes from the fact that for $|U|\gg 1$ the wavefunction 
for the light fermion is given by the superposition of two orbitals that are peaked
at the positions of the heavy particles. When the latter are not nearest neighbor, corresponding to $n>1$, the two orbitals have vanishing overlap as $|U|$ becomes large.
As a consequence, the light fermion cannot easily delocalize between the two sites 
implying that the binding mechanism is less robust.

\section{Conclusions}
In conclusion, we have presented a detailed discussion of  three-body bound states of 
interacting fermions in one-dimensional optical lattice. These states only occur when
the tunneling rates of the two spin components are different.
 As a consequence the asymmetric model (\ref{aHubb})  exhibits a new strongly correlated 
phase at low but finite density, corresponding to a Luttinger liquid of trimers, 
which is completely absent in the usual attractive Hubbard model. 
This happens when the densities of the two components are commensurate allowing binding of all atoms into stable trimers. Then the
interacting gas of trimers form a one-component Luttinger liquid. 
Thanks to the Pauli exclusion principle, these phases are expected to be particularly stable 
against three-body recombination and therefore experimentally accessible with ultra-cold Fermi gases. 

  Our results provide one more example of the importance of few-body physics to identify
the fundamental degrees of freedom of a full many-body system.

\label{sec5}





\begin{thebibliography}{99}

\bibitem{RMP2008}
For a review of both theoretical and experimental status see,
\textit{e.g.}, I. Bloch, J. Dalibard, and W. Zwerger,
Rev. Mod. Phys. \textbf{80}  (2008) 885; S.~Giorgini, L. P.~Pitaevskii, and S.~Stringari,
Rev. Mod. Phys. \textbf{80}  (2008) 1215.

\bibitem{repulsivelybound} K. Winkler, G. Thalhammer, F. Lang, R. Grimm, J. Hecker Denschlag, 
A. J. Daley, A. Kantian, H. P. Buechler, P. Zoller, Nature 441 (2006) 853. 

\bibitem{efimov} S. Knoop, F. Ferlaino, M. Mark, M. Berninger, H. Schoebel, H.-C. Naegerl, 
R. Grimm, Nature Physics 5   (2009) 227.

\bibitem{fedichev} P.O. Fedichev, M.J. Bijlsma, and P. Zoller,  Phys. Rev. Lett. \textbf{92}  (2004) 080401.
\bibitem{Tc2005} G. Orso and G.V. Shlyapnikov, Phys. Rev. Lett. \textbf{95} (2005) 260402. 

\bibitem{esslinger} H. Moritz, T. St\" oferle, K. G\" unter, M. K\" ohl, and T. Esslinger,
Phys. Rev. Lett. \textbf{94} (2005) 210401. 

\bibitem{Hulet2010}  Y. Liao, A.S.C. Rittner, T. Paprotta, W. Li, G.B. Partridge,  
R.G. Hulet, S.K. Baur, E.J. Mueller, Nature \textbf{467} (2010) 567. 

\bibitem{Orso2007}
G.~Orso,
Phys. Rev. Lett \textbf{98} (2007) 070402.

\bibitem{Drummond2007}
H.~Hu,  X.-J.~Liu, and P. D.~Drummond,
Phys. Rev. Lett. \textbf{98} (2007) 070403.


\bibitem{FFLO}
P.~Fulde and R. A.~Ferrell,
Phys. Rev. \textbf{135} (1964) A550;
A. I.~Larkin and Yu. N.~Ovchinikov,
Sov. Phys. JETP \textbf{20} (1965) 762.

\bibitem{yang} K. Yang, Phys. Rev. B \textbf{63} (2001) 140511.



\bibitem{Feiguin}
A.~Feiguin and F.~Heidrich-Meisner, Phys. Rev. B \textbf{76} (2007) 220508(R);  
G. G. Batrouni, M. H. Huntley, V. G. Rousseau, and R. T. Scalettar, 
Phys. Rev. Lett. \textbf{100} (2008) 116405;
M. Rizzi, M. Polini, M.A. Cazalilla, M.R. Bakhtiari, M.P. Tosi, and R. Fazio, 
Phys. Rev.B \textbf{77} (2008) 245105; 
M. Tezuka and M. Ueda, 
Phys. Rev. Lett. \textbf{100} (2008) 110403;
F. Heidrich-Meisner, G. Orso, A. Feiguin, Phys. Rev. A. 81 (2010) 053602.
 
 \bibitem{bosonization}
E. Burovski, G. Orso, and Th. Jolicoeur, Phys. Rev. Lett. \textbf{103} (2009) 215301.

 \bibitem{PRL2010} G. Orso, E. Burovski, and Th. Jolicoeur,
Phys. Rev. Lett. \textbf{104} (2010) 065301.

\bibitem{GiamarchiCazalillaHo2005}
M. A.~Cazalilla, A. F.~Ho, and Th.~Giamarchi,
Phys. Rev. Lett. \textbf{95} (2005) 226402.

\bibitem{Batrouni2009}
G. G.~Batrouni, M.J.~Wolak, F.~Hebert, and V.G.~Rousseau,
Europhys. Lett. \textbf{86}  (2009) 47006.

\bibitem{DasSarma2009}
B.~Wang, Han-Dong Chen, and S.~Das~Sarma, 
Phys. Rev. A \textbf{79} (2009) 051604(R).

 
\bibitem{Wille2008}
E.~Wille, F.M. Spiegelhalder, G. Kerner, D. Naik, A. Trenkwalder, G. Hendl, F. Schreck, 
R. Grimm, T.G. Tiecke, J.T.M. Walraven, S.J.J.M.F. Kokkelmans, E. Tiesinga, and P.S. Julienne, 
Phys. Rev. Lett. \textbf{100} (2008) 053201.

\bibitem{dieckmann}
A.-C. Voigt, M. Taglieber, L. Costa, T. Aoki, W. Wieser, T. W. H�nsch, and K. Dieckmann, 
Phys. Rev. Lett. \textbf{102} (2009) 020405.

\bibitem{unequal}
G. Orso, L. P. Pitaevskii, and S. Stringari,
Phys. Rev. A \textbf{77} (2008) 033611.

\bibitem{takahashi}
M. Takahashi, Progr. Theor. Phys. \textbf{43} (1970) 917.

\bibitem{trion}
J. G. Groshaus \textit{et al.}, Phys. Rev. Lett. \textbf{98} (2007) 156803.

\bibitem{P}D. S. Petrov, Phys. Rev. A \textbf{67}  (2003) 010703(R).

\bibitem{KML}O. I. Kartavtsev and A. V. Malykh, JETP Letters \textbf{86} (2007) 625.

\bibitem{KM}O. I. Kartavtsev, A. V. Malykh, J. Phys. B: At. Mol. Opt. Phys. \textbf{40}  
(2007) 1429.

\bibitem{NT}Y. Nishida and S. Tan, Phys. Rev. Lett. \textbf{101}  (2008) 170401.

\bibitem{LTWP}J. Levinsen, T. Tiecke, J.Walraven, and D. Petrov, Phys. Rev. Lett. \textbf{103} (2009) 153202.

\bibitem{KMS}O. I. Kartavtsev, A. V. Malykh, and S. A. Sofianos, JETP \textbf{108}  (2009) 365.

\bibitem{HHP}K. Helfrich, H.-W. Hammer, D. Petrov, Phys. Rev. A \textbf{81} (2010) 042715.

\bibitem{PP}L. Pricoupenko and P. Pedri, Phys. Rev. A \textbf{82} (2010)  033625.



\bibitem{mattis} D. C. Mattis, Rev. Mod. Phys. \textbf{58} (1986) 361.

\bibitem{valiente10} 
M.~Valiente, D.~Petrosyan, A.~Saenz, Phys. Rev. A \textbf{81} (2010) 011601(R). 

\bibitem{keilmann09}
T.~Keilmann, I.~Cirac, T.~Roscilde, Phys. Rev. Lett. \textbf{102}
  (2009) 255304.

\bibitem{azaria09}
P.~Azaria, S.~Capponi, P.~Lecheminant, Phys. Rev. A \textbf{80} (2009) 041604; 
S. Capponi, G. Roux, P. Lecheminant, P. Azaria, E. Boulat, S.R. White, Phys. Rev. A \textbf{77} (2008)  013624.


\bibitem{kantian09}
A.~Kantian, M.~Dalmonte, S.~Diehl, W.~Hofstetter, P.~Zoller, A.J. Daley, Phys.
  Rev. Lett. \textbf{103} (2009) 240401.

\bibitem{luescher10}
A.~L\"uscher, A.~L\"auchli,   arXiv:0906.0768 (unpublished).

\bibitem{noteP} The total \textsl{momentum}, in contrast,  is not conserved because
 the \textsl{continuous} traslational symmetry is broken by the lattice. This causes the appearance of Umklapp collisions, see for instance G. Orso, L. P. Pitaevskii, and S. Stringari, Phys. Rev. Lett. \textbf{93} (2004) 020404. 

\bibitem{twobody} G. Orso, L. P. Pitaevskii, S. Stringari, and M. Wouters,
Phys. Rev. Lett. \textbf{95} (2005) 060402; M. Wouters and G. Orso, Phys. Rev. A \textbf{73} (2006) 012707.


\bibitem{moelmer} R. T. Piil, N. Nygaard, and K. M\" olmer, Phys. Rev. A \textbf{78} (2008) 033611.



















\end{thebibliography}
\end{document}